\documentclass[pre,reprint,aps,superscriptaddress,floatfix]{revtex4}
\usepackage{longtable}

\usepackage[utf8]{inputenc}
\usepackage[T1]{fontenc}
\usepackage{mathptmx}

\usepackage{graphicx}
\usepackage{dcolumn}
\usepackage{color}
\usepackage{bm}
\graphicspath{{.},{./Ps/},{../Ps/},{../../Ps/}}
\begin{document}

\title{  Harmonic and subharmonic waves on the surface of a vibrated liquid drop}


\author{Ivan S. Maksymov}
\affiliation{Centre for Micro-Photonics, Swinburne University of Technology, Hawthorn, Victoria, 3122, Australia}
\author{Andrey Pototsky}
\affiliation{Department of Mathematics, Faculty of Science Engineering and Technology, Swinburne University of Technology, Hawthorn, Victoria, 3122, Australia}
\email{apototskyy@swin.edu.au}

\begin{abstract}
Liquid drops and vibrations are ubiquitous in both everyday life and technology, and their combination can often result in fascinating physical phenomena opening up intriguing opportunities for practical applications in biology, medicine, chemistry and photonics. Here we study, theoretically and experimentally, the response of pancake-shaped liquid drops supported by a solid plate that vertically vibrates at a single, low acoustic range frequency. When the vibration amplitudes are small, the primary response of the drop is harmonic at the frequency of the vibration. However, as the amplitude increases, the half-frequency subharmonic Faraday waves are excited parametrically on the drop surface.
  We develop a simple hydrodynamic model of a one-dimensional liquid drop to analytically determine the amplitudes of the harmonic and the first superharmonic components of the linear response of the drop. In the nonlinear regime, our numerical analysis reveals an intriguing cascade of instabilities leading to the onset of subharmonic Faraday waves, their modulation instability and chaotic regimes with broadband power spectra. We show that the nonlinear response is highly sensitive to the ratio of the drop size and Faraday wavelength. The primary bifurcation of the harmonic waves is shown to be dominated by a period-doubling bifurcation, when the drop height is comparable with the width of the viscous boundary layer. Experimental results conducted using low-viscosity ethanol and high-viscocity canola oil drops vibrated at $70$\,Hz are in qualitative agreement with the predictions of our modelling.

\end{abstract}

\maketitle

\section{Introduction}
Parametrically excited waves on the surface of a vertically vibrated fluid, originally observed in $1831$ by Faraday \cite{Faraday}, have become a paradigmatic example of a nonlinear wave system that exhibits highly complex dynamics, including periodic \cite{Benjamin_54}, quasiperiodic \cite{henderson_miles_1990,miles_1984,jiang_1996} and chaotic behaviour \cite{Punzmann09,Xia12,Shats2010,Shats2012}. Several recent studies have also opened a series of new frontiers for potential applications of Faraday waves going well beyond fluid dynamics and nowadays including photonics \cite{Tarasov_2016, Huang_17}, metamaterials \cite{Domino_2016}, alternative sources of energy \cite{Alazemi_2017}, and biology \cite{Sheldrake_2017}.

Because Faraday waves can readily be observed in a vertically vibrated container filled with fluid, a large and growing body of experimental research uses them to investigate nonlinear wave phenomena such as rouge surface waves and solitons \cite{Shats2010, Xia12}. Faraday waves emerge on the flat fluid surface as standing subharmonic waves that oscillate at half of the vibration frequency. The nonlinear dynamics regime of the emerging waves is in particular rich and very sensitive to the aspect ratio of the container, fluid depth, vibration frequency and amplitude, as well as to the presence of surfactants \cite{henderson_miles_1991,Punzmann09,Xia12,Shats2010,Shats2012}. The fundamental instability of the standing waves is associated with the modulation instability of their amplitude and it was experimentally observed in cylindrical containers with small and large aspect ratios \cite{henderson_miles_1990,jiang_1996,Punzmann09}. Modulation of the amplitude occurs on a much larger times scale than the vibration period and can be detected by observing the development of zero frequency sideband in the power spectrum of optical signal reflected from the fluid surface. 
The importance of the boundary and contact line effects on the stability threshold of standing waves has been recognized in early experiments performed in cylindrical and square tanks with small aspect ratio \cite{henderson_miles_1990,jiang_1996}. For example, it was shown that viscous dissipation at the contact line increases wave damping close to solid boundaries \cite{Benjamin_54,miles_1984}.

In a different class of experiments, {various} types of external forcing, including mechanical vibration, electrowetting and surface acoustic waves, have been used to excite capillary waves on the surface of a liquid drop. A partially wetting drop in contact with a solid plate exhibits a discrete spectrum of eigenfrequencies and vibration modes that strongly depend on the wettability of the solid and mobility of the contact line  \cite{Bostwick1,Chang_2013}. When a drop is periodically forced, its linear response can be cast into the framework of a standard damped-driven harmonic oscillator \cite{Bostwick3,Chang_2015}, which implies that the longtime response of the drop is harmonic at the frequency of the forcing. For example, a harmonic nature of the primary response was confirmed in experiments using $10\dots 100$ $\mu$l liquid drops driven by surface acoustic waves \cite{Yeo_2010,Baudoin_2012}. 

At a sufficiently strong vibration, the deformation of free surface enters a nonlinear regime and therefore the temporal signature of the drop response loses a simple periodic character. While general characteristics of the nonlinear response strongly depend on the driving frequency, in a particular experiment with $\sim 10$ $\mu$l water drops, the low frequency ($\sim 100$ Hz) modulation of a $20$ MHz acoustic wave was shown to induce harmonic, superharmonic and subharmonic excitations of volumetric oscillation modes \cite{Baudoin_2012}. In the cited paper, the subharmonic response was observed at the modulation frequency that is approximately twice as large as the Rayleigh volumetric frequency of the drop. However, when a drop of a similar volume was driven by a high-frequency acoustic wave without modulation, the subharmonic half-frequency component was noticeably absent in the nonlinear response \cite{Yeo_2010}. Instead, the power spectrum of surface deformation developed a broadband component at an approximately $100$ times lower frequency than the driving one.  

The volumetric oscillation frequency $f_V$ of a sessile drop decreases with its volume $V$. Note that the classical Rayleigh-Lamb result for levitated drops gives the $f_V\sim V^{-1/2}$ dependence. In the view of the early experiments \cite{Baudoin_2012}, it is apparent that, for larger drops (volume of $100$ $\mu$l or more) driven at around $100$ Hz, the volumetric oscillation modes should not be excited. Indeed, in this regime a lens-shaped drop undergoes a much more dramatic shape transformation resulting in horizontal elongation and formation of a worm-like structure \cite{Hemmerle_2015, Pucci_2011,Pucci_2013}. In the elongated state, subharmonic Faraday waves develop on the surface of the drops but interactions with moving contact lines and nonlinearity of surface deformation renders a highly complex spatio-temporal wave dynamics. Thus far, a fully nonlinear theory of the Faraday waves that takes into account boundary and contact line effects is missing.


In this present paper we study, theoretically and experimentally, instabilities of harmonic waves emerging on the surface of a pancake-shaped liquid drop located on top of a vertically vibrated solid plate. Unlike in fluids constrained by the side walls of a container, a liquid drop changes its shape in responses to vibration and therefore can be used as a prototype of a domain with soft adjustable boundaries. 
A weak external vibration linearly excites harmonic capillary waves on the drop surface \cite{Bostwick3,Chang_2015}. To excite subharmonic Faraday waves oscillating at half the driving frequency, a certain threshold amplitude has to be passed. In the following, we will describe the transition from the harmonic waves to subharmonic Faraday waves in a pancake-shaped drop and study their secondary bifurcations as a function of a gradually changing amplitude of vertical vibrations.
 
To describe the dynamics of a pancake-shaped drop with a sufficiently large volume, we adopt a simplified version of the reduced long-wave model \cite{Pototsky_2018}. When the driving frequency is in the $10\dots100$\,Hz range, the resonance due to volumetric oscillation modes is not observed. We develop a linear response theory of a one-dimensional drop that takes into account oscillations of the meniscus at the drop edge. Excess Laplace pressure in the meniscus excites harmonic waves on the surface of the drop, whose amplitude is shown to obey a driven-damped Mathieu equation. Using the small amplitude expansion, with the driving amplitude as a small parameter, we derive the amplitudes of harmonic and first superharmonic responses. 

Then, we numerically integrate the full nonlinear model equations to study the instabilities of the harmonic waves. By matching the model parameters to describe the dynamics of an ethanol drop vibrated at $20\dots 30$\,Hz frequencies, we gradually increase the vibration amplitude to observe the onset of the subharmonic Faraday waves via super or sub-critical period-doubling, or torus bifurcations of the harmonic wave solutions. The bifurcation scenario is shown to be highly sensitive to the relationship between the horizontal drop size to the Faraday wavelength. Moreover, when the viscosity of the fluid is  increased several-folds while keeping all other parameters unchanged, we find that the primary instability of the harmonic waves is dominated by the supercritical period-doubling bifurcation. In the nonlinear regime, longer Faraday waves interact with shorter harmonic waves, thereby giving rise to a complex mixed state whose temporal signature is characterised by the presence of subharmonic, harmonic and superharmonic peaks in the temporal power spectrum of the surface deformations. When the vibration amplitude is further increased, we observe modulation instability of the Faraday waves that occurs on the time scale much larger than the vibration period. Stronger vibration gives rise to a chaotic response with broadband power spectrum across all frequencies.

To validate our findings, we conduct a series of experiments with a low-viscosity pancake-like ethanol drop and a high-viscosity canola oil drop. In both series of experiments, the Teflon plate supporting the drops is vibrated at $70$\,Hz. The response of the drops to the vertical vibration is recorded using laser light reflected from the drop surface. We show that our experimental results are in a qualitative agreement with the predictions of the long wave model.

\section{Hydrodynamic model of a vibrated liquid drop}
Correct description of the contact line motion is one of the biggest hurdles in modelling the dynamics of liquid drops that rest on a solid plate. To avoid the well-known hydrodynamic singularities at the true contact line \cite{Huh_Scriven71}, we use the standard regularization method based on the molecularly thin precursor film that emanates from the foot of the drop and covers the entire solid plate \cite{deGennes}. The equilibrium contact angle is determined by balancing the pressure in the precursor film and in the drop. For small contact angles, the total pressure can be written in terms of the drop thickness $h(x,y)$ \cite{deGennes}
\begin{eqnarray}
  \label{eq1}
P=-\sigma(\partial_{xx}+\partial_{yy}) h +\rho g h - \Pi(h),
\end{eqnarray}
where $\sigma$ and $\rho$ are the surface tension and the fluid density and $\Pi(h)$ is the disjoining pressure that describes the long-range van der Waals forces giving rise to the formation of the precursor film. Any stationary drop profile $h(x,y)$ can be found from Eq.\,(\ref{eq1}) by setting $P=C$, where the constant $C$ is determined by the volume of the fluid. Generally, the function $\Pi(h)$ that allows the existence of a steady drop with non-zero contact angle is given by \cite{Schwartz1998}
\begin{eqnarray}
  \label{dp}
\Pi(h)= \frac{B}{h_{\infty}^n} \left(\frac{h_{\infty}^n}{h^n}-\frac{h_{\infty}^m}{h^m}\right),
\end{eqnarray}
where $B$ is some constant, $h_{\infty}$ is the precursor film thickness and the $(m,n)$ are some integers. Without any loss of generality of the results, we follow \cite{uwe10,Pototsky_2018} and choose $n=6$ and $m=3$. With this choice, the disjoining pressure Eq.\,(\ref{dp}) can be written in terms of the Hamaker constant $A_H$
\begin{eqnarray}
  \label{dp1}
\Pi(h)= \frac{A_H}{h^3}\left(\frac{h_{\infty}^3}{h^3}-1\right).
\end{eqnarray}
The equilibrium contact angle $\theta \ll 1$ is related to $A_H$ and $h_\infty$ via \cite{Schwartz1998}
\begin{eqnarray}
  \label{dp2}
A_H\approx  \frac{5h_\infty^2}{3}\sigma \theta^2.
\end{eqnarray}

A reduced hydrodynamic model describing the dynamics of liquid drop an non-zero Reynolds numbers under the action of external vertical vibration was developed in Ref.~\cite{Pototsky_2018}. Here, we adopt a simplified version of the model \cite{Pototsky_2018} that captures all essential features of the flow field, but has a more simple expression for the nonlinear terms. Thus, we only take into account the long-wave deformations of the liquid-gas interface and assume a quadratic dependence of the horizontal fluid velocity ${\bm u}(x,y,z)$ on the vertical coordinate $z$
\begin{eqnarray}
  \label{eq2}
{\bm u}={\bm \Phi}\left(\frac{z^2}{2}-hz\right),
\end{eqnarray}
that satisfies the boundary conditions at the solid plate ${\bm u}(z=0)=0$ and the liquid-gas interface $\partial_z{\bm u}(z=h)=0$ for some arbitrary function ${\bm \Phi}(x,y)$. The flow field across the layer ${\bm q}=\int_0^h {\bm u}\,dz$ can be expressed in terms of ${\bm \Phi}$
\begin{eqnarray}
  \label{eq3}
{\bm q}=-{\bm \Phi}\frac{h^3}{3}.
\end{eqnarray}
The Navier-Stokes equation in the long-wave approximation is
\begin{eqnarray}
  \label{eq4}
\rho\left(\partial_t {\bm u} + {\bm \nabla}({\bm u}^2)+\partial_z({\bm u}w)\right)=\mu\partial_{zz}{\bm u}-{\bm \nabla} P,
\end{eqnarray}
where $\mu$ is the dynamic viscosity, $w$ is the vertical component of the velocity and $P$ is the pressure.
Integrating Eq.\,(\ref{eq4}) over $z$ and using the kinematic boundary condition $\partial_t h +({\bm u}\cdot {\bm \nabla} h) = w$, we obtain similar to \cite{best2013,bes13a}
\begin{eqnarray}
  \label{eq5}
  \rho\left[\partial_t {\bm q}+\frac{6}{5}{\bm \nabla} \cdot \left(\frac{{\bm q}\otimes {\bm q}}{h}\right)\right]&=&-\frac{3\mu {\bm q}}{h^2}-h{\bm \nabla} P,\nonumber\\
  \partial_t h + ({\bm \nabla}\cdot {\bm q}) &=&0,
  \end{eqnarray}
where ${\bm q}\otimes {\bm q}$ denotes the matrix product.

In the case of a drop supported by a solid plate that vibrates vertically with the amplitude $A_0$ and frequency $\Omega$, Eqs.\,(\ref{eq5}) are valid in the co-moving frame of the plate. The pressure $P$ is taken from Eq.\,(\ref{eq1}) with $g$ replaced by $g(1+a\cos{\Omega t})$, where $a=A_0\Omega^2/g$ is the dimensionless vibration amplitude. We note that the validity of system Eqs.\,(\ref{eq5}) is restricted to small contact angles and long-wave deformations of the drop height $h$. In addition, the characteristic horizontal deformation wave length $\lambda$ must be larger than the length of the viscous boundary layer $l=\sqrt{2\mu/(\rho\Omega)}$ associated with the vibration frequency $\Omega$ \cite{BP16}.

For a flat film of thickness $h_0 \gg h_{\infty}$ the disjoining pressure term $\Pi(h)$ can be neglected. Linearizing Eqs.\,(\ref{eq5}) about the trivial steady state $h=h_0$ and ${\bm q}=0$ we obtain the damped Mathieu equation
\begin{eqnarray}
  \label{eq6}
 \partial_{tt} \tilde{h} +\frac{3\mu}{\rho h_0^2}\partial_t \tilde{h} + h_0\left(\frac{\sigma}{\rho} \Delta^2+g(1+a\cos{\Omega t})\Delta\right) \tilde{h}=0,
  \end{eqnarray}
where $\tilde{h}$ denotes small deviation of the film thickness from $h_0$.

\begin{figure}[ht]
\centering
\includegraphics[width=0.8\textwidth]{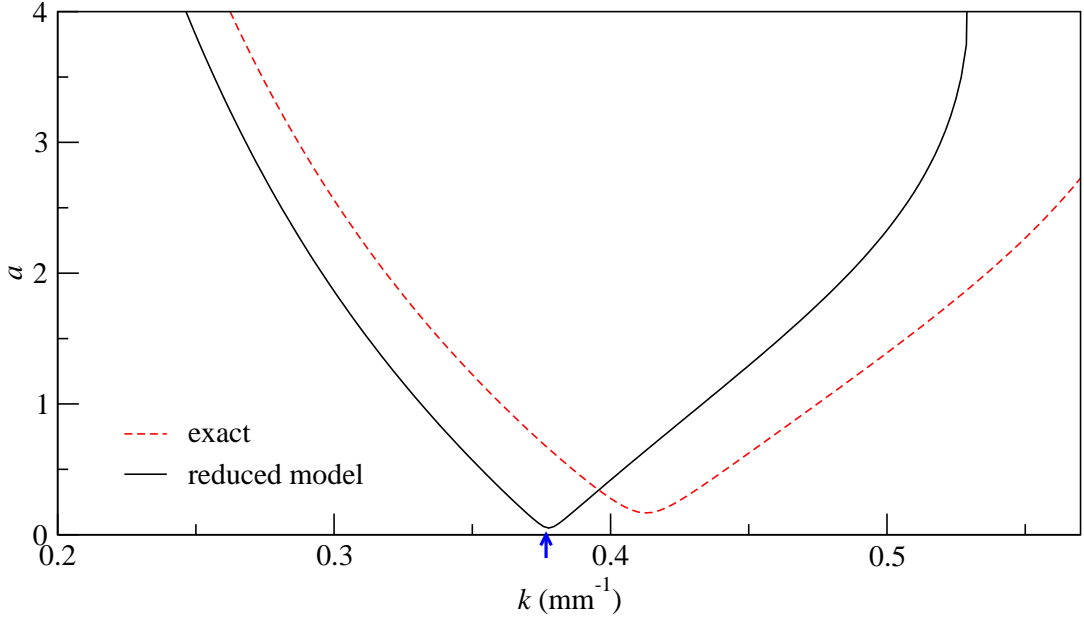}
\caption{(Color online)  Stability diagram for the subharmonic Faraday waves onset in $2$ mm thick ethanol film with $\rho=789$ kg/m$^3$, $\sigma=0.022$ N/m, $\mu=0.0012$ Pa$\cdot$s, vibrated at $20$\,Hz. The solid line is obtained from the reduced model Eq.\,(\ref{eq6}), the dashed line corresponds to the exact stability threshold \cite{Kumar96}. The arrow indicates the critical wave vector from Eq.\,(\ref{eq8}).
  \label{F1}}
\end{figure}

For inviscid and undriven fluids ($\mu=a=0$), Eq.\,(\ref{eq6}) gives the well-known dispersion relation of plane waves $\tilde{h}\sim e^{i\omega t +i{\bm k}\cdot {\bm r}}$ on the surface of the flat film $\omega^2=\left(\frac{\sigma}{\rho} k^3+gk\right)\tanh{kh_0}$ \cite{Kumar96}
\begin{eqnarray}
  \label{eq7}
 \omega^2= h_0\left(\frac{\sigma}{\rho} k^4+gk^2\right) \approx \left(\frac{\sigma}{\rho} k^3+gk\right)\tanh{kh_0}.
  \end{eqnarray}
To further validate our model, we calculate the stability diagram for the onset of the Faraday waves of wave number $k$ on the surface of a flat film with thickness $h_0\gg h_{\infty}$. The exact stability threshold found according to \cite{Kumar96} is compared with the prediction of the reduced model Eq.\,(\ref{eq6}) in Fig.\,\ref{F1} for the case of a $2$ mm thick ethanol film vibrated at $20$ Hz. For these parameter values, the agreement is reasonable for the first sub-harmonic tongue. We note, however, that the model deviates from the exact theory if the vibrating frequency, or the film thickness $h_0$, is further increased.

At a low value of the kinematic viscosity $\mu/\rho$, the critical vibrational amplitude $a$ corresponding to the tip of the tongue in Fig.\,\ref{F1} is small. This allows estimating the onset wave number $k_c$ of the Faraday waves, which are excited subharmonically at half of the driving frequency and harmonically at the frequency of the forcing. Using the standard result for the classical undamped Mathieu equation \cite{Mathieu}, from Eq.\,(\ref{eq6}) in the limit of $\mu/\rho \rightarrow 0$ we obtain
\begin{eqnarray}
  \label{eq8}
 2k_c^2=\sqrt{(g\rho/\sigma)^2+\rho(n\Omega)^2/(\sigma h_0)}-g\rho/\sigma,
  \end{eqnarray}
where odd (even) integers $n=1,2,3,\dots$ correspond to the sub-harmonic and harmonic waves, respectively.
The critical wave vector $k_c$ calculated from Eq.\,(\ref{eq8}) for the first subharmonic $(n=1)$ tongue is shown by the arrow in Fig.\,\ref{F1}. 
\section{Vibrated pancake shaped drop}
Next we nondimensionalize Eqs.\,(\ref{eq5}) by scaling the time with $2/\Omega$, the coordinates $(x,y)$ with $k_c^{-1}$, where $k_c$ from Eq.\,(\ref{eq8}) corresponds to the tip of the first sub-harmonic tongue, and the film thickness with $h_0$ and the fluid flux ${\bm q}$ with $h_0 \Omega/ (2k_c)$. In what follows we associate the dimensionless wave vector $k=1$ with the Faraday wave vector and the corresponding wavelength $2\pi$ with the Faraday wavelength $\lambda_F=2\pi$.

The dimensionless dynamic equations of the drop are obtained from Eqs.\,(\ref{eq5}) by setting $\rho=1$, $3\mu=\gamma=6\mu/(\rho h_0^2\Omega)$ and replacing the pressure term Eq.\,(\ref{eq1}) by
\begin{eqnarray}
  \label{eq9}
 P=-\Gamma\Delta h + Gh(1+a\cos(2t))+H\,h^{-3}(1-(\beta/h)^3),
\end{eqnarray}
with the dimensionless $\Gamma=4\sigma h_0 k_c^4/(\rho\Omega^2)$, $G=4gh_0k_c^2/\Omega^2$, $H=4A_H k_c^2/(\rho h_0^3\Omega^2)$ and $\beta=(h_\infty/h_0)$. Note that $\Gamma+G=1$ and we use the same notations for the scaled variables as for the original physical variables. The scaled damping parameter $\gamma=6\mu/(\rho h_0^2\Omega)$ can be written in terms of the ratio of the viscous boundary layer length $l=\sqrt{2\mu/(\rho \Omega)}$ and the drop height $h_0$, namely $\gamma=3(l/h_0)^2$. 
The scaled disjoining pressure parameters $H$ and $\beta$  must be chosen to be sufficiently small. Specifically, we choose $\beta \ll 1$ and $H\ll G$ so that the disjoining pressure in a flat film with height $h=1$ can be safely neglected as compared with the hydrostatic pressure. Additionally, we require that $h=1$ corresponds to the equilibrium dimensionless height of a steady pancake shaped drop. Note that $h_0$ is related to the equilibrium contact angle $\theta$ via $\rho gh_0^2=2\sigma(1-\cos(\theta))$. Consequently, for any given fluid parameters $\rho$ and $\sigma$, a fixed value of $h_0$ corresponds to a constant value of the contact angle $\theta$. This implies that in the chosen here scaling (when the dimensionless drop height is $h=1$), the dimensionless Hamaker constant $H$ and $\beta$ are not independent. In fact, $H\sim \beta^2$ as follows from Eq.\,(\ref{dp2}).
In our numerical simulations, we choose $\beta$ as an independent parameter and determine for any given $G$ the corresponding value of $H$ from the Maxwell equal-area construction to the function $f(h)=Gh+H\,h^{-3}(1-(\beta/h)^3)$, as explained in Ref.\,\cite{Pototsky_2018}. In what follows we use $\beta=0.05$, which is of the similar order of magnitude as in Ref.\,\cite{Schwartz1998} where the disjoining pressure model Eq.\,(\ref{dp}) was validated for liquid drops spreading on solid surfaces.

For the sake of simplicity, we consider a one-dimensional version of Eqs.\,(\ref{eq5}) with the drop profile $h(x,t)$ and fluid flux $q(x,t)$. Following \cite{Pototsky_2018}, we numerically solve Eqs.\,(\ref{eq5}) in a periodic domain $x\in [-L,L]$ using a semi-implicit spectral method with the time step $\Delta t \approx 10^{-3}$ and the spatial discretization step $\Delta x$ of the order of $\beta$. The initial condition is given by a stationary drop, which is obtained by solving $-\Gamma \partial_{xx} h + Gh + H\,h^{-3}(1-(\beta/h)^3)=C$, where the constant $C$ controls the volume of the drop.
The system size $2L$ and the horizontal width of the pancake shaped drop $W$ are chosen significantly larger than the Faraday wave length $\lambda_F=2\pi$, i.e.  $2L,W \gg 2\pi$. Additionally, in order to avoid self-interactions due to periodic boundaries, we require $2L-W \gg 2\pi$.  

For ethanol drop with parameters as in Fig.\,\ref{F1}, vibrated at $20$ Hz, we obtain $\Gamma=0.3$, $G=0.7$ and $\gamma=0.018$. In the dimensionless units, the drop with the horizontal width $W\approx 48.5$, which is vibrated at $a=0.1$ is shown in Fig.\,\ref{F2}(a). It rests on the precursor film of thickness $\beta=0.05$, the domain length is $2L=24\pi$.  At $a=0.1$ the amplitude of harmonic waves excited on the drop surface is two orders of magnitude smaller than the drop height.

We note that currently available computational power limits the values of the precursor film parameter to $\beta \sim 10^{-2}\dots10^{-1}$. Indeed, to guarantee the stability of the numerical scheme with the spatial discretization step $\Delta x$ that is of the order of $\beta$ \cite{Schwartz1998}, $N=10^3\dots 10^4$ equidistant discretization points in the domain of the length $2L=24\pi$ is required. Since our aim is to detect possible bifurcations of the vibrated drop, the Eqs.\,(\ref{eq5}) must be integrated over a sufficiently large interval of time, at least much larger than the vibration period equal to $\pi$ time units. We integrate Eqs.\,(\ref{eq5}) with the spatial resolution of $N=5000$ over $2000$ vibration periods, which yields the average simulation time of $\approx 20$ min for one set of parameters on a $3.3$\,GHz Intel workstation. We remark that at this resolution, the numerical simulation of the two-dimensional Eqs.\,(\ref{eq5}) would require unaffordable computer resources and prohibitively long simulation time. 

\subsection{Harmonic waves}
\label{harm_waves}
\begin{figure}[ht]
\centering
\includegraphics[width=0.8\textwidth]{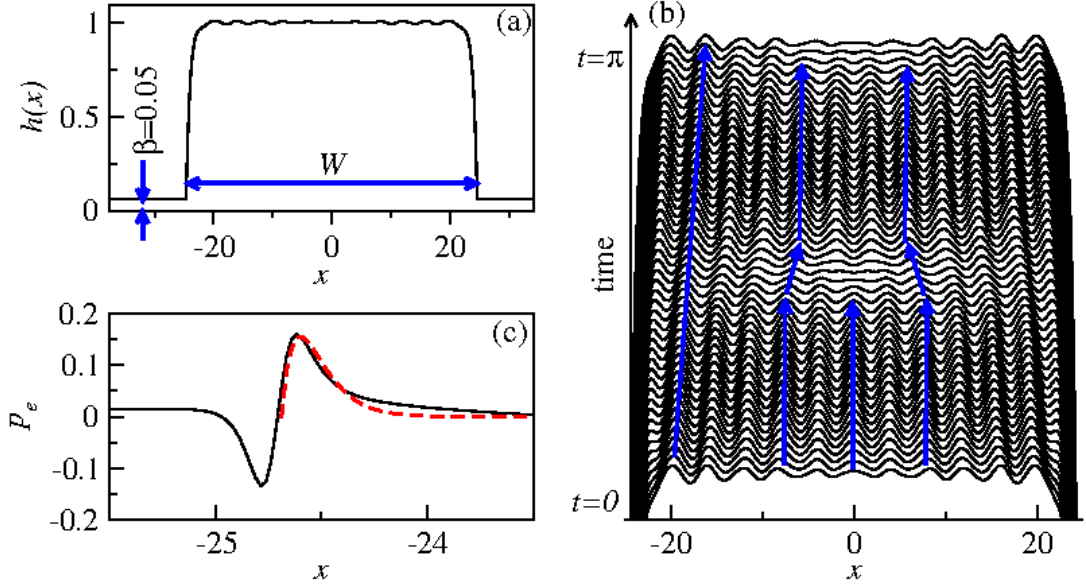}
\caption{(Color online) (a) Profile of the drop with $\beta=0.05$ and horizontal size $W\approx 48.5$, vibrated with amplitude $a=0.1$. Domain size is $2L=24\pi$.
(b) Temporal evolution of the harmonic waves on the surface of the drop over one period of the vibration $T=\pi$. The arrows indicate the propagating and standing waves. (c) Excess Laplace pressure $p_e=-\Gamma \partial_{xx} h(x,t)+\Gamma \pi^{-1}\int_{t}^{t+\pi} \partial_{xx} h(x,t)\,dt$ for the drop in (a). The dashed line in (c) is obtained by fitting the function $\sim (x-x_0)e^{(x-x_0)/b}$.
  \label{F2}}
\end{figure}

The response of a pancake-shaped drop to external vibration is qualitatively different from the response of an infinitely extended flat film because of the presence of the oscillating meniscus at the drop edge. Thus, at infinitesimally small vibration amplitudes $a \ll 1$, the linear response of the drop is always harmonic with the meniscus, oscillating at the forcing frequency $\Omega$.
In Fig.\,\ref{F2}(a), we show the profile of a weakly vibrated one-dimensional drop at $a=0.1$. The amplitude of the harmonic waves on the drop surface remains of the order of $10^{-2}$. The temporal evolution of the waves over one vibration period $T=\pi$ is shown in Fig.\,\ref{F2}(b). It is important to emphasize that these waves are not simple standing waves excited at the left and right edge of the drop. Instead, we see that close to the drop edges, the waves propagate towards the drop centre, as indicated by the arrows in Fig.\,\ref{F2}(b). However, for the symmetric oscillations, the single wave in the centre of the drop is a indeed a standing wave. 

The mechanism of the harmonic waves excitation can be understood by studying the Laplace pressure generated by the meniscus. More specifically, we compute the excess Laplace pressure $p_e(x,t)$, determined as the difference between the time-dependent pressure $-\Gamma \partial_{xx} h(x,t)$ and the time-averaged pressure $-\Gamma T^{-1}\int_{t}^{t+T} \partial_{xx} h(x,t)\,dt$, where we average over one oscillation period $T=\pi$. Fig.\,\ref{F2}(c) shows a snapshot of $p_e$ around the left edge of the drop from Fig.\,\ref{F2}(a). The time-dependent $p_e(t)$ can be obtained by multiplying the function in Fig.\,\ref{F2}(c) by the factor $\cos{(2t+\phi)}$, where $\phi$ is a phase shift between the forcing $\cos{2 t}$ and the meniscus oscillations. Note that the function $p_e(x,t)$ has two maxima: the left maximum excites surface waves in the precursor film, but the right maximum excites harmonic waves in the drop. Let $x=\pm x_0$ be the positions of the left and the right edges of the drop. In the following we focus on the maximum of the function $p_e(x)$ responsible for the oscillations of the drop. At any moment of time, 
the total excess pressure, generated by the left and the right oscillating drop edges, can be well approximated by
\begin{eqnarray}
  \label{eq10}
  p_e=aF\cos{(2 t+\phi)}\left[ (x-x_0)e^{(x-x_0)/b} - (x+x_0)e^{-(x+x_0)/b} \right],
 \end{eqnarray}
where $b$ is of the order of the meniscus width, and $F$ is some scaling factor that describes the strength of the response. 

As a next step, we derive an effective evolution equation for the symmetric small amplitude surface deformations of the drop. We extend the pressure term in the dimensionless version of Eqs.\,(\ref{eq5}) by $p_e$ from Eq.\,(\ref{eq10}) and restrict all possible solutions to the interval within the drop, i.e. $x\in [-x_0,x_0]$.  Linearizing Eqs.\,(\ref{eq5}) in the limit of weak vibration $(a\ll 1)$ about the trivial steady state  $h=1$ and $q=0$, for the deformation of the drop surface $\tilde{h} \ll 1$ we obtain
\begin{eqnarray}
  \label{eq11}
  \partial_{tt} \tilde{h} + \gamma \partial_t \tilde{h} + (\Gamma \partial_{x}^4  - G(1+a\cos(2t))\partial_{x}^2 )\tilde{h} - \partial_{xx}p_e&=&0.
\end{eqnarray}
We are looking for symmetric oscillations of the drop and consider only such perturbations of the trivial steady state that are $2x_0$-periodic. Under these assumptions we apply the discrete Fourier transforms $\tilde{h}(x,t)=\sum_{k=1}^{\infty} \hat{h}_k\cos(\pi k x/x_0)$ and reduce  Eq.\,(\ref{eq11}) to a non-homogeneous damped Mathieu equation for the amplitudes $\hat{h}_k$, where $\partial_{xx}p_e$ plays the role of an external drive
\begin{eqnarray}
  \label{eq12}
  \partial_{tt} \hat{h}_k + \gamma \partial_t \hat{h}_k + \left[\Gamma \left(\frac{\pi k}{x_0}\right)^4  + G(1+a\cos(2t)) \left(\frac{\pi k}{x_0}\right)^2 \right]\hat{h}_k &=& af_k\cos{(2t+\phi)},
\end{eqnarray}
with $f_k$ given by the cos-Fourier transform of the even function $\partial_{xx} \left[ (x-x_0)e^{(x-x_0)/b} - (x+x_0)e^{-(x+x_0)/b} \right]$.

It is noteworthy that the additional driving term $af_k\cos{(2t+\phi)}$ in Eq.\,(\ref{eq12}) does not change the stability of its solution \cite{Mathieu}. Therefore, in the linear regime, the Faraday instability threshold for the waves on the surface of a pancake-shaped drop remains unchanged as compared with an infinitely extended flat film. However, this conclusion only applies approximately in the limit of a large drop volume, when the nonlinear effects at the meniscus have little effect on the dynamics of waves at the centre of the drop. Generally, in finite volume drops, the onset of the Faraday instability is associated with a period-doubling or a torus bifurcation point. The critical value of the vibration amplitude, at which the Faraday waves are excited, depends on the drop volume \cite{Pototsky_2018}.
\subsection{Linear response and small amplitude expansion }
\label{linear_resp}
\begin{figure}[ht]
\centering
\includegraphics[width=0.8\textwidth]{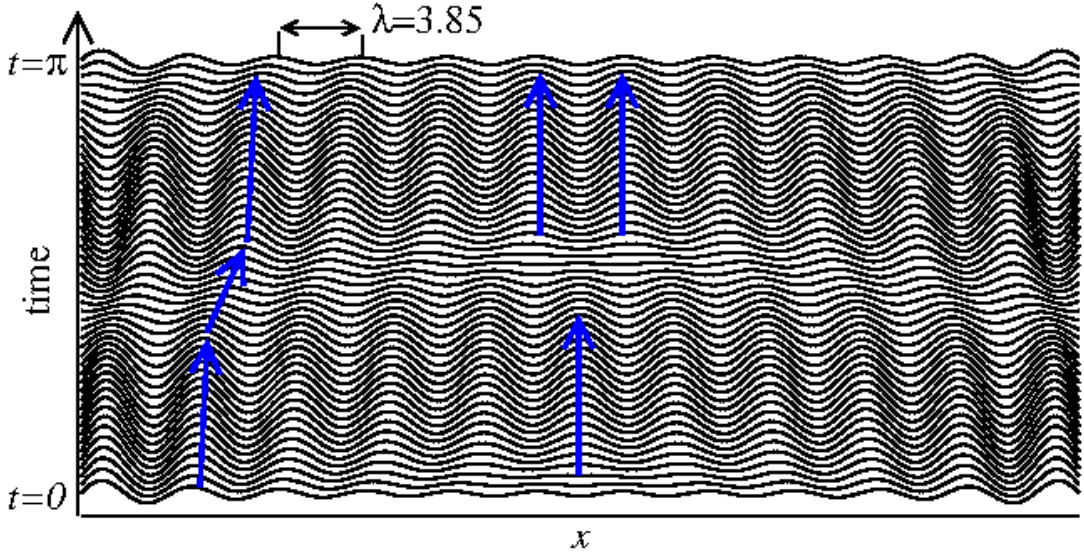}
\caption{(Color online) Temporal evolution of the deformation of the drop surface, computed from Eq.\,(\ref{eq15}) over one vibration period $T=\pi$ with parameters as in Fig.\,\ref{F2}.
  \label{F3}}
\end{figure}

For vibrational amplitudes $a$ well below the Faraday threshold in Fig.\,\ref{F1}, the response of the drop can be determined from Eq.\,(\ref{eq12}) analytically. We assume $a \ll 1$ and seek the solution in the form $\hat{h}_k=a(\hat{h}^{(0)} + a \hat{h}^{(1)} + a^2\hat{h}^{(2)} +\dots)$, where the primary response $\hat{h}^{(0)}$ is anticipated linear in $a$. At the order $a^1$ we obtain 
\begin{eqnarray}
  \label{eq14}
  \partial_{tt} \hat{h}^{(0)} + \gamma \partial_t \hat{h}^{(0)} + \omega^2 \hat{h}_0 = f_k\cos{(2t+\phi)},
\end{eqnarray}
with $\omega^2=\left[\Gamma \left(\frac{\pi k}{x_0}\right)^4  + G \left(\frac{\pi k}{x_0}\right)^2 \right]$. Eq.\,(\ref{eq14}) is an equation of a periodically forced damped harmonic oscillator with natural frequency $\omega$ so that its long time solution is given by
\begin{eqnarray}
  \label{eq15}
  \hat{h}^{(0)} =\frac{f_k\cos\left(2t + \psi \right)}{\sqrt{(4-\omega^2)^2+4\gamma^2}},
\end{eqnarray}
where  $\psi = \phi -\arctan{[2\gamma/(\omega^2-4)]}$.
The evolution of the deformation of the drop surface, calculated from Eq.\,(\ref{eq15}), over one vibration period is shown in Fig.\,\ref{F3}.

The linear response theory correctly predicts the visible wavelength of the surface deformation. Thus, the dominant wave length $\lambda=2x_0/k\approx 3.85$ in Fig.\ref{F3} corresponds to the resonant peak in Eq.\,(\ref{eq15}), i.e. the integer $k$ is so selected that the term $(\omega^2(k)-4)^2$ has a minimal possible value. For comparison, the visible wavelength in Fig.\ref{F2}(b) is $\lambda\approx 3.9$.  In addition, Eq.\,(\ref{eq15}) correctly describes the main features of the wave dynamics close to the drop centre (compare with Fig.\,\ref{F2}(b)). 

At the next order $a^2$, we have
\begin{eqnarray}
  \label{eq16}
  \partial_{tt} \hat{h}^{(1)} + \gamma \partial_t \hat{h}^{(1)} + \omega^2 \hat{h}^{(1)} - f^{(1)}(e^{2it}+e^{-2it})\left(e^{2it+i\psi}+e^{-2it-i\psi}\right)&=&0,
\end{eqnarray}
with $f^{(1)}=\frac{f_kGk^2}{4\sqrt{(4-\omega^2)^2+4\gamma^2}}$.
The long time solution of Eq.\,(\ref{eq16}) is
\begin{eqnarray}
  \label{eq17}
  \hat{h}^{(1)} =\frac{2f^{(1)}\cos{\psi}}{\omega^2}+\frac{2f^{(1)}\cos\left(4t + \psi_1 \right)}{\sqrt{(16-\omega^2)^2+16\gamma^2}},
\end{eqnarray}
with $\psi_1=\psi-\arctan{[4\gamma/(\omega^2-16)]}$.

According to Eq.\,(\ref{eq17}), the response of the drop at order $a^2$ contains the second harmonic signal $\cos{4t}$ and a constant shift $\frac{2f^{(1)}\cos{\psi}}{\omega^2}$. 
We remark that Eq.\,(\ref{eq17}) explains the presence of the superharmonic $\sim \cos{4t}$ component in the response of drop driven by surface acoustic waves \cite{Yeo_2010}. Thus, the appearance of this component is a linear effect, associated with the deviation of the temporal response function from the simple $\sin$-shaped function of time. 
%
\section{Onset of the Faraday waves and the mixed state}
\label{bif}
%
 To study nonlinear drop response, we employ the following naive continuation method. Eqs.\,(\ref{eq5}) are solved numerically for gradually varying (increasing or decreasing) values of the vibration amplitude $a$: $a_n=n\Delta$, $(n=0,1,2,3,\dots)$, where the step $\Delta$ is fixed. For each $n$ Eqs.\,(\ref{eq5}) are integrated over the time interval, equivalent to $1000$ vibration cycles. The initial conditions for the $n$-th run are taken as the solution from the previous $(n-1)$-th run.  To artificially recreate the common experimental conditions, we add a small-amplitude white noise $\xi(x,t)$, with $\langle \xi(x,t)\xi(x',t')\rangle=D\delta(x-x')\delta(t-t')$ to the second equation in Eqs.\,(\ref{eq5}), where $D$ denotes the noise strength $D\sim 10^{-4}$.
The deformation $\delta h(t)$ of the drop surface at $x=0$ and its power spectrum $S_f$ are used to characterise the temporal response. The spatial Fourier transformation of the drop height is averaged over $1000$ vibration cycles to obtain the average spatial spectrum  $S_k=T^{-1}\int_0^T|\hat{h}_t(k)|^2\,dt$, where $\hat{h}_t(k)$ stands for the discrete Fourier transform of the field $h(x,t)$ taken from the interval $x\in[-L/2,L/2]$. 
We present the response of a fixed-volume ethanol drop as in Fig.\,\ref{F2} vibrated at different frequencies in the range between $f=20$ Hz and $f=28$ Hz.
\subsection{Drop size to Faraday wavelength ratio}
\begin{figure}[ht]
\centering
\includegraphics[width=0.8\textwidth]{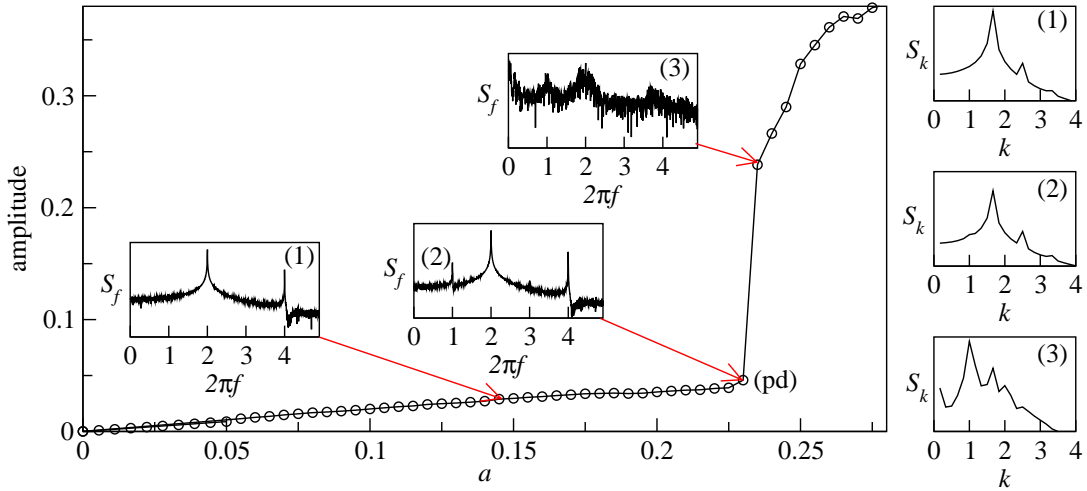}
\caption{(Color online) Average response amplitude of the ethanol drop from Fig.\,\ref{F2}, vibrated at $f=20$ Hz, as a function of $a$. The amplitude $a$ is gradually increased with the step $\Delta=0.005$. The harmonic waves lose their stability via a subcritical period-doubling bifurcation (pd) at $a=0.23$. the insets labelled as (1,2,3) show the temporal and spatial spectra $S_f$ and $S_k$ in the logarithmic scale for the selected values of $a$ indicated by the arrows. 
  \label{F4}}
\end{figure}

The average amplitude of $\delta h(t)$ in the units of the drop height $h_0$ is shown as a function of $a$ in Fig.\,\ref{F4} for a fixed-volume ethanol drop vibrated at $f=20$ Hz. For $a<0.23$, the response amplitude increases linearly with $a$. In accord with the results of the previous section, at small vibration amplitudes the response is harmonic, as evidenced by the temporal spectrum $S_f$ that has a major peak at the driving frequency $2\pi f=2$ and the second smaller peak at the second harmonic $2\pi f=4$ (label (1)). The spatial spectrum shows the dominance of a single mode with the wave vector $k\approx 1.7$.
At a certain critical amplitude $a=0.23$, the drop undergoes a subcritical period-doubling bifurcation, as confirmed by a jump of the response amplitude and by the appearance of the first sub-harmonic component $2\pi f =1$ in the power spectrum (label (2)).  The drop enters a new mixed highly irregular (chaotic) state, characterised by the broad band temporal spectrum (label (3)). The spatial power spectrum in the mixed state is dominated by the Faraday waves with the wave vector $k=1$.

\begin{figure}[ht]
\centering
\includegraphics[width=0.8\textwidth]{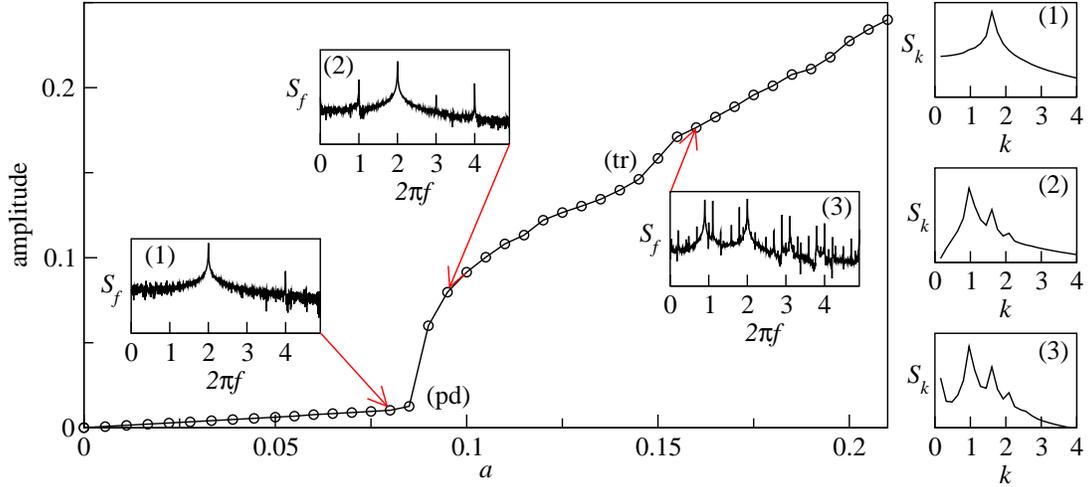}
\caption{(Color online) Average response amplitude of the ethanol drop, vibrated at $f=21$\,Hz, as a function of $a$. Continuation by gradually increasing $a$ with step $\Delta=0.005$. Harmonic waves lose their stability via supercritical period-doubling bifurcation (pd) at $a=0.08$. Modulation instability sets in via a torus bifurcation (tr) at $a=0.15$. Insets labeled (1,2,3) show the temporal and spatial spectra $S_f$ and $S_k$ in the logarithmic scale for selected values of $a$, as indicated by arrows. 
  \label{F5}}
\end{figure}

Significantly, the characteristics of the drop response are highly sensitive to the relationship between the Faraday wavelength $\lambda_F$ and the horizontal size of the drop $W$. To better fit the experimental conditions, where the drop volume is fixed, we vary the vibrational frequency $f$ and thus change the wavelength of the Faraday waves in the dimensional units. Accordingly, the dimensionless parameters $\Gamma$, $G$ and $H$ vary with $f$. Simultaneously, we vary $W$ in such a way that the dimensional horizontal size of the drop is kept fixed. Thus increasing the vibration frequency from $f=20$ Hz to $f=21$ Hz, which corresponds to $\approx 4\%$ shorter Faraday waves, we observe a dramatic change in the bifurcation scenario, as shown in Fig.\,\ref{F5}. Contrary to Fig.\,\ref{F4}, the primary bifurcation of the harmonic waves is a supercritical period-doubling bifurcation that occurs at a significantly smaller amplitude $a=0.08$. The newly born mixed state has a high degree of the temporal order, as characterized by the sharp subharmonic and harmonic peaks in the temporal spectrum (label (2)). Similar to Fig.\,\ref{F4}, the mixed state is dominated by the Faraday wave vector $k=1$, as shown by the spatial spectrum (label (2)). 

At around $a=0.15$ the mixed state undergoes a secondary bifurcation, which corresponds to the modulation instability of the wave amplitude. From the point of view of the dynamical systems theory, this instability corresponds to the torus bifurcation (point tr), as confirmed by the appearance of a small frequency peak in the temporal spectrum (label (3)). 

\begin{figure}[ht]
\centering
\includegraphics[width=0.8\textwidth]{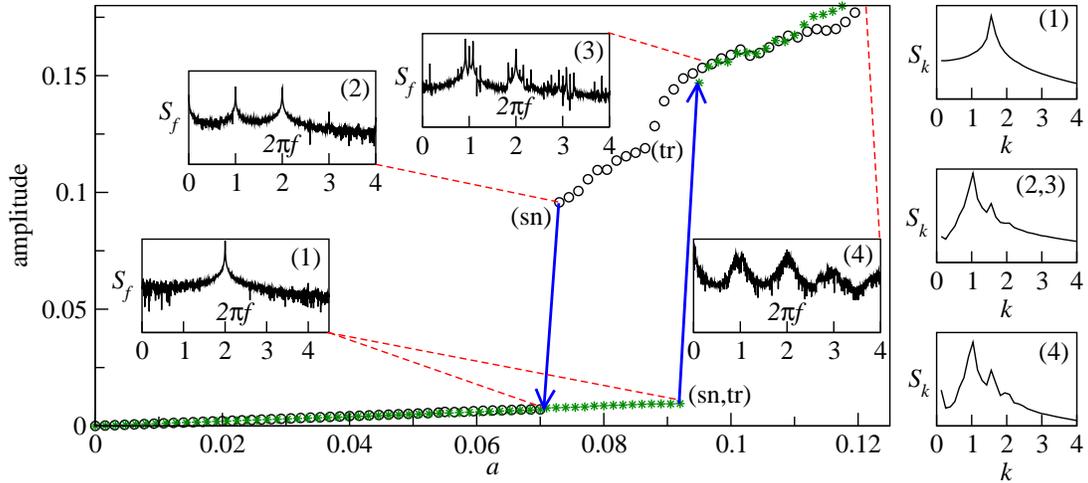}
\caption{(Color online) Average response amplitude of the ethanol drop, vibrated at $f=28$ Hz, as a function of $a$.  Gradually increasing (asterisk) and gradually decreasing (circle) $a$ with step $\Delta=0.0015$. sn and tr correspond to the saddle-node and torus bifurcations, respectively. The insets labeled (1,2,3,4) show the temporal and spatial spectra $S_f$ and $S_k$ in the logarithmic scale for the selected values of $a$ indicated by the dashed lines. 
  \label{F6}}
\end{figure}

By scanning the other values $f$ we find that at $f=28$\,Hz, the primary instability of the harmonic waves is most likely a torus bifurcation, as shown in Fig.\ref{F6}. When $a$ is gradually increased with the step $\Delta=0.0015$, the harmonic waves follow the "asterisk" branch and lose their stability at around $a=0.092$ (label (sn,tr)). The response amplitude jumps significantly and any detectable subharmonic peak in the temporal spectrum is absent (label (1)), which points towards either a torus, or a saddle-node bifurcation. The new state corresponds to the modulated Faraday waves, which is confirmed by the presence of the frequency sidebands in the temporal spectrum (label (2)). Spatial spectrum of the mixed state has a strong peak at $k=1$ (label (2,3)). Further increase of $a$ leads to the onset of a chaotic temporal response at around $a=0.12$ (label (4)).

When the vibration amplitude $a$ is gradually decreased from $a=0.12$ with the step $\Delta=0.0015$, the response amplitude follows the "circle" branch that stretches to $a=0.07$. At $a=0.087$ (label (tr)), the modulated Faraday waves are replaced by the non-modulated standing waves via a reversed torus bifurcation, as confirmed by the temporal spectrum (label (2)). Interestingly, however, the spatial signature of the modulated and standing waves is very similar, as shown by the spatial spectrum (label (2,3) ). For $a<0.07$ (label (sn)), the branch of the standing non-modulated Faraday waves does not exist and the drop response is harmonic. The subcriticality of bifurcations gives rise to the hysteresis loop, as indicated by the thick blue arrows in Fig.\ref{F6}. In the range of $a$ between $a=0.07$ and $a=0.092$, multiple responses are possible -- either small amplitude harmonic waves with dominant wave vector $k=1.7$ or standing or modulated Faraday waves of larger amplitude with the dominant wave vector $k=1$.
%
\begin{figure}[ht]
\centering
\includegraphics[width=0.8\textwidth]{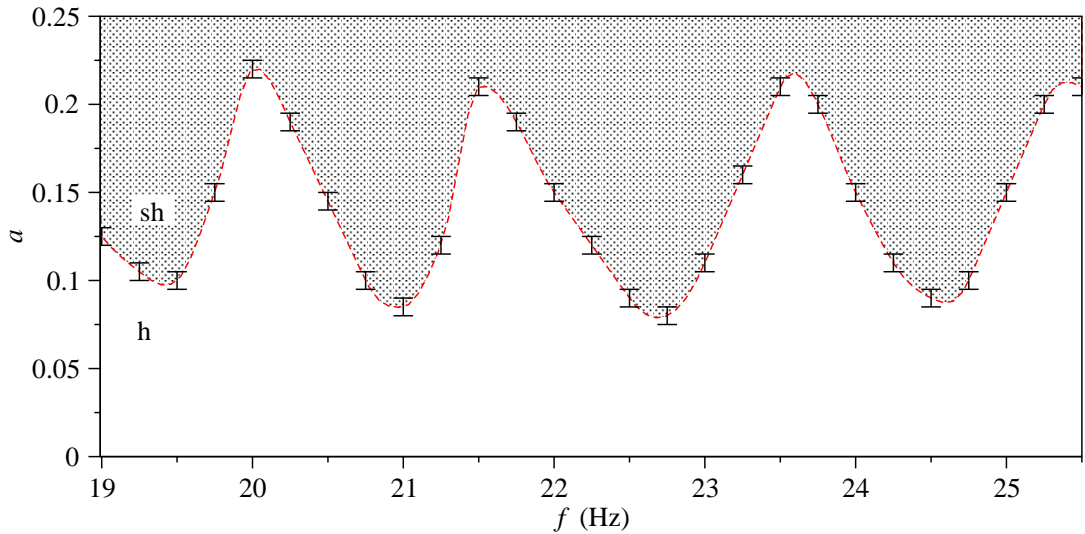}
\caption{(Color online)   Frequency-amplitude phase diagram of the harmonic (h) and subharmonic (sh) responses for a fixed volume ethanol drop. The response is subharmonic in the shaded area. The boundary between the harmonic and subharmonic regions is obtained by gradually increasing the vibration amplitude $a$ by step $0.01$ (size of the error bars), when the frequency $f$ is kept fixed. 
  \label{F7}}
\end{figure}
A chaotic response is observed at large amplitudes $a>0.12$ and this is characterised by the broad band temporal spectrum (label (4)).
The broadening of spectral peaks in case of the Faraday waves excited in liquid films has been the focus of several experimental studies in the past \cite{Punzmann09,Xia12}. In the cited papers, the appearance of the triangular peaks was linked to the modulation of the amplitudes of the Faraday waves that occurs on a much longer time scale than the vibration period. Our model confirms those experimental findings and explains the observed spectra from the point of view of the bifurcation theory. In particular, the amplitude modulation of the Faraday waves results from a torus bifurcation and the spectral broadening is most likely to be due to the destruction of the invariant torus via the collision with a homoclinic orbit. 

High sensitivity of the drop response to the ratio between Faraday wave length and the horizontal drop size is visualized in Fig.\,\ref{F7}, where we plot the boundary between the harmonic and the subharmonic responses in the $(a,f)$ plane for a fixed volume ethanol drop. The boundary is obtained by gradually increasing the vibration amplitude $a$, when the frequency $f$ is kept fixed. For comparison, when changing the frequency from $f=20$ Hz to $f=25$ Hz, the Faraday wave length decreases by 20\%. Such a high sensitivity towards the frequency $f$ can be explained by a simple geometric commensurability condition. Namely, if the horizontal drop size is an integer multiple of the half of the Faraday wavelength $\lambda_F/2=\pi$, then subharmonic Faraday waves are more easily excited on the drop surface. Indeed, at $f=19.5$ Hz approximately $15$ half wavelengths $\lambda_F/2$ (in dimensionless units) fit onto the surface of the drop. This corresponds to the tip of the first tongue in Fig.\,\ref{F7}. The tips of the second, third and fourth tongues correspond to $f\approx 21$ Hz, $f\approx 22.5$ Hz and $f\approx 24.5$ Hz, when approximately $16$, $17$ and $18$ half wavelengths $\lambda_F/2$ fit onto the surface of the drop, respectively. 
\subsection{Changing the viscosity}
\begin{figure}[ht]
\centering
\includegraphics[width=0.8\textwidth]{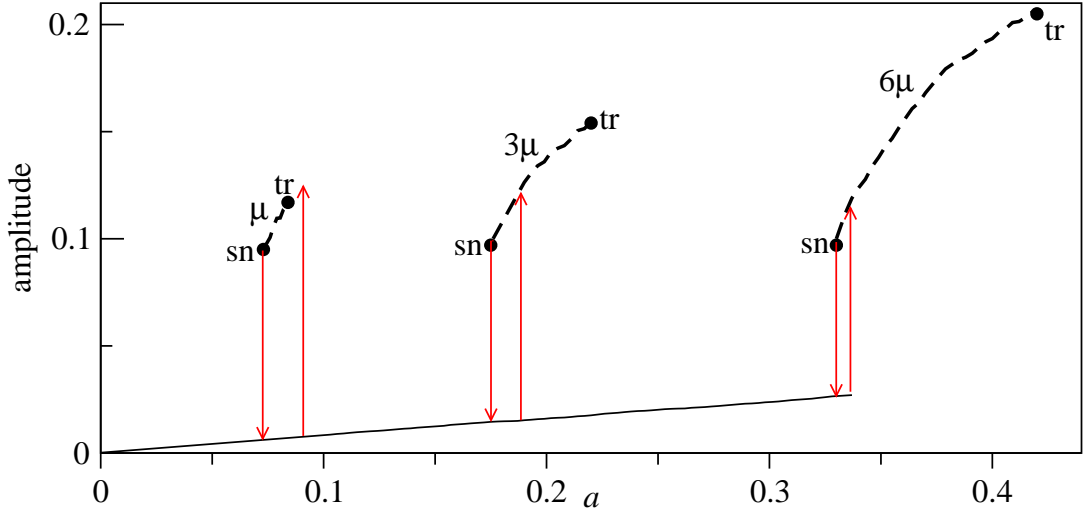}
\caption{Increasing the viscosity: (dashed line) part of the subharmonic branch between the sadle-node (sn) and the torus (tr) bifurcation points for three different values of the viscosity: $\mu$ , $3\mu$ and $6\mu$, where $\mu$ is the viscosity of the ethanol. The vibration frequency is $f=28$ (Hz) and other parameters as in Fig.\,\ref{F6}. The arrows indicate points of transitions from the harmonic (solid line) to subharmonic (dashed line), when $a$ is gradually increased or decreased.
  \label{F8}}
\end{figure}

We predict that in highly viscous fluids, in which the viscous boundary layer $\sqrt{2\mu/(\rho \Omega)}$ is comparable with the drop height $h_0$, the modulation instability of the subharmonic Faraday waves is pushed towards larger values of the vibration amplitude.  In this case the primary instability of the harmonic waves is dominated by the period-doubling bifurcation and the modulation sidebands are not found in the temporal spectrum. This regime is characterised by the dimensionless damping parameter $\gamma=6\mu/(\rho h_0^2\Omega)$ of the order of $10^{-1}\dots 10^0$.  Thus, when $\mu$ is increased, the torus bifurcation point (tr) on the subharmonic branch occurs at progressively larger values of the vibration amplitude $a$, as shown in Fig.\,\ref{F8} for a liquid that is three times ($3\mu$) and six times ($6\mu$) more viscous than ethanol. Vertical arrows in Fig.\,\ref{F8} indicate the hysteresis region, i.e. the points of transition from the harmonic to subharmonic regimes (or backwards), when $a$ is gradually increased (or decreased). For comparison, the part of the subharmonic branch for the ethanol from Fig.\,\ref{F6} between the saddle-node and the torus bifurcation points is added to Fig.\,\ref{F8}. 
Similar to a liquid film, the critical vibration amplitude that marks the onset of the Faraday waves increases as the viscosity of the drop is increased. At large values of $\mu$, the primary bifurcation of the harmonic waves is a sub-critical period-doubling bifurcation, as indicated by the significant jump in the response amplitude. The region of multistability, where subharmonic and harmonic responses coexist, significantly narrows as $\mu$ is increased six-folds. It is plausible to assume that further increase of $\mu$ will result in a super-critical period-doubling bifurcation of the harmonic waves. 

The critical role of the viscosity in the onset of the modulation instability was recognized in the early works on the weakly nonlinear theory of the Faraday waves in cylindrical containers \cite{miles_1984}. Nonlinear waves on the surface of inviscid fluids can be described using the Lagrangian formalism that leads to a Hamiltonian system for slowly varying amplitude of the waves. The simplest possible quadratic nonlinearity in the Lagrangian equations of motion, gives rise to periodic solutions for the slowly varying amplitude, which explains the modulation instability of the Faraday waves. However, an arbitrary weak linear damping completely changes the phase space of the Hamiltonian system, by destroying the periodic orbits. Therefore, in the weakly nonlinear regime, the modulation instability, which is associated with the torus bifurcation point on the subharmonic branch, does not exist. Our numerical results demonstrate the importance of strong nonlinearities for the onset of the modulation instability in viscous fluids. In agreement with weakly nonlinear theories, our model shows that in highly viscous fluids, the modulation of the Faraday waves does not occur at weak vibration.

\section{Experimental results}

In this section, we experimentally demonstrate that the primary response of a liquid drop to low-amplitude vertical vibrations is harmonic at the frequency of the forcing, but an increase in the vibration amplitude results in the excitation of half-frequency subharmonic surface Faraday waves. In the first set of measurements, we investigate the response of a low-viscosity ethanol drop to demonstrate that an increase in the vibration amplitude first leads to the onset of Faraday waves and then results in sharp amplitude modulation sidebands observed in the frequency spectra due to a secondary bifurcation. In the second set of measurements, we study the response of a high-viscosity \textit{canola oil} drop. (The viscosity of canola oil is approximately $50$ times that of ethanol). We observe the onset of Faraday waves at much higher vibration amplitudes as compared with the ethanol drop and we confirm that in the case of high viscosity of the constituent liquid the frequency spectra have well-defined peaks with no signs of amplitude modulation. 

\begin{figure}[ht]
\centering
\includegraphics[width=0.5\textwidth]{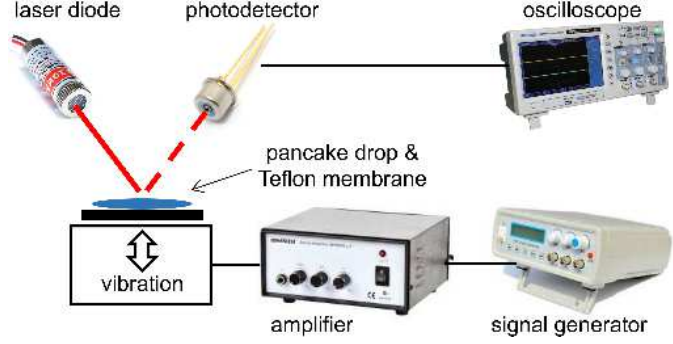}
\caption{(Color online) Schematic of the experimental setup. A subwoofer covered by a thin Teflon plate is used as the source of vertical vibration. The sinusoidal vibration signal of frequency $f=70$\,Hz is synthesises with a digital signal generator and amplified with an audio amplifier. The response of the liquid drop sitting on top of the Teflon plate is measured by using a continuous wave red laser diode and a photodetector. The detected signals are visualised with an oscilloscope and sent to a laptop for post-processing.
\label{Setup}}
\end{figure}

To measure the temporal response of liquid drops subjected to vertical vibration, we developed a setup (Fig.\,\ref{Setup}) consisting of a non-collimated red laser diode (Besram Technology, China, $650$\,nm wavelength, $1$\,mW power) and a photodetector (Adafruit, USA). The intensity of light reflected from the liquid is modulated due to the vertical vibration of the drop and the onset of Faraday waves on the liquid surface. We record these signals with Audacity software and Fourier-transform them with Octave software to obtain frequency spectra of surface deformation.

As a source of vertical vibration, we use a subwoofer (Yamaha SW-P330, Japan, $85$\,W, $30...200$\,Hz frequency response) covered by an opaque Teflon plate. The subwoofer is driven by a pure sinusoidal signal of frequency $f=70$\,Hz generated by a digital signal generator. The operation at this particular frequency provides a viable opportunity for qualitative comparison of theoretical and experimental results, also allowing us to efficiently control the shape of the vibrated drops and reduce noise in experimental traces. We verified that the addition of the plate does not change the frequency response of the subwoofer \cite{Erz11}. At low vibration amplitudes, we could optically detect weak, not perceivable to human ear signals with the twice AC mains frequency ($100$\,Hz), which were removed from experimental spectra as part of routine post-processing.    

Purified canola oil and $95\%$ v/v ethanol were purchased from Coles, Australia. The viscosity of canola oil equals approximately $0.057$\,Pa$\cdot$s ($\sim 47$ times the viscosity of ethanol). Drops with an approximately $2$\,mm thickness were created on top of the Teflon plate by using a syringe. In accord with the exact linear theory \cite{Kumar96}, at the vertical vibration frequency of $70$\,Hz the Faraday wave length for $2$-mm-thick canola oil and ethanol films is $6.1$\,mm and $5.7$\,mm, respectively. Thus, the size of the drops was chosen to be approximately $10$ times larger than the predicted wavelengths, also avoiding any contact between the drops and the edges of the Teflon plate. 

\begin{figure}[ht]
\centering
\includegraphics[width=0.95\textwidth]{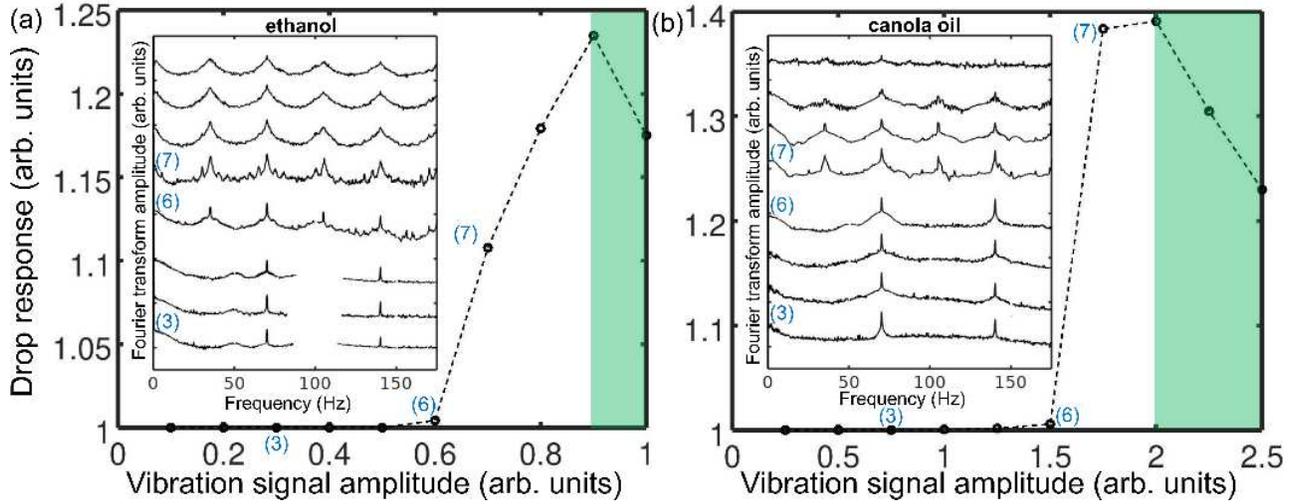}
\caption{(Color online) (a) Experimental average response of the \textit{ethanol} drop subjected to vertical vibration at $70$\,Hz frequency plotted as a function of the vibration amplitude. The inset shows the power spectra obtained by Fourier-transforming the measured time-domain signals. The labels in parenthesis correlate the spectra to the experimental points in the main panel. (b) Experimental average response of the \textit{canola oil} drop subjected to vertical vibration at $70$\,Hz frequency plotted as a function of the vibration amplitude. Note the difference between the maximum vibration amplitude in (a) and (b). Also note the presence of the modulation sidebands (the spectrum label $7$ for the ethanol drop) and their absence in the spectra of the \textit{canola oil} drop. The shaded regions in the main panels correspond to the regime of chaotic oscillations resulting in strong diffuse scattering leading to a decrease in the optical intensity of the detected signal.
\label{Experiment1}}
\end{figure}

In agreement with the theoretical picture in Fig.\,\ref{F5}, in Fig.\,\ref{Experiment1}(a) we observe that at low vibration amplitudes the spectral response of the \textit{ethanol} drop is dominated by the peaks at the vibration frequency $2\pi f = 2$ and its second harmonic $2\pi f = 4$ ($2\pi f = 2$ corresponds to $f=70$\,Hz). The harmonic waves lose their stability via supercritical period-doubling bifurcation as evidenced by the appearance of the peaks $2\pi f = 1$ and $2\pi f = 3$. Significantly, our measurements show the onset of the modulation instability after the generation of sub-harmonic peaks, observed as sharp modulation sidebands around the peaks in the frequency spectra label ($7$).

When the vibration amplitude is further increased, we observe that the amplitude modulation results in considerable broadening of the spectral peaks, which assume a typical triangular shape previously reported for liquid films subjected to vibration \cite{Punzmann09,Xia12}. Finally, at an even higher amplitude, the drop response becomes noisy, which is explained by the transition to a chaotic behaviour predicted by our theory.

In measurements of the the \textit{canola oil} drop [Fig.\,\ref{Experiment1}(b)], we observe the period-doubling bifuraction point at a significantly higher vibration amplitude as compared with the ethanol droplet, which agrees with the theoretical result in Fig.\,\ref{F8}. Significantly, our measurements reproduce the theoretically prediction of no amplitude modulation. Indeed, one can see the appearance of the sub-harmonic peaks at $2\pi f = 1$ and $2\pi f = 3$ without any trace of sidebands around the peaks.

\begin{figure}[ht]
\centering
\includegraphics[width=0.4\textwidth]{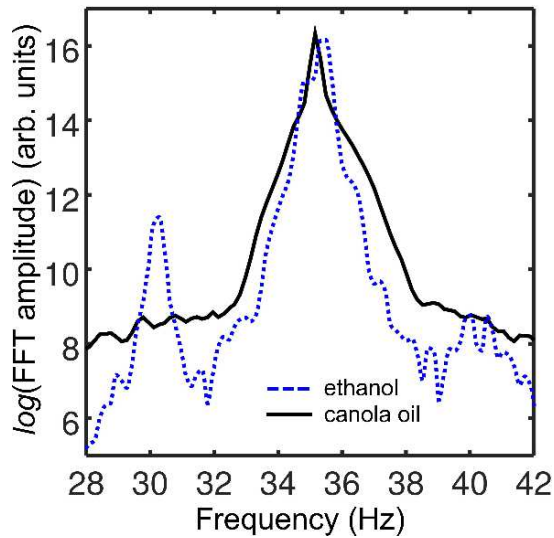}
\caption{(Color online) Closeups of the sub-harmonic peak at $2\pi f = 1$ corresponding to $f = 35$\,Hz in the experiment. The modulation sidebands can be seen in the spectrum of the ethanol drop (dashed line), but they are absent in the spectrum of the \textit{canola oil} drop (solid line).
\label{Experiment2}}
\end{figure}

The difference between the spectra with the traces of modulation sidebands (ethanol) and without them (canola oil) can be better observed in Fig.\,\ref{Experiment2}, which shows a closeup of the sub-harmonic peaks at $2\pi f = 1$ corresponding to $35$\,Hz in the framework of our experiment. The frequency resolution of the spectra is approximately $0.1$\,Hz. One can see that the peak for the \textit{canola oil} drop has a well-defined lineshape, but that for the ethanol drop appears to consist of sub-peaks and also has two side peaks offset by approximately $\pm 5$\,Hz. In agreement with the previous works \cite{Punzmann09,Xia12}, as the vibration amplitude is increased, these sub-peaks contribute to the evolution of the narrow main peaks into broad triangular spectral features, which can be seen in the inset in Fig.\,\ref{Experiment1}(a).   

\section{Conclusion}

We have investigated, experimentally and theoretically, the response of a flattened pancake-shaped liquid drop to external vibrations at the frequencies much higher than the frequency of the volumetric oscillations. We have showed that in this regime harmonic oscillations of the meniscus at the contact line generate time-periodic excess Laplace pressure that excites capillary ripples on the surface of the drop. When the vibration amplitude is small, the amplitude of the ripples remains small compared with the drop height, which allows us to linearize the hydrodynamic equations about the steady flat liquid-gas interface and derive a driven damped Mathieu equation for the surface wave amplitudes. By expanding the surface deformation into a power series of the small driving amplitude $a^n$, ($n=1,2,\dots$), we derived a closed analytical expression for the harmonic (linear $\sim a^1$,) and superharmonic (quadratic $\sim a^2$) responses. In the second order $\sim a^2$, we find a time-independent response component associated with the change in the average drop height. The latter, in conjunction with the conservation of volume, gives rise to horizontal elongation of the drop.

We numerically solve the fully-nonlinear reduced-model equations to reveal complex primary and secondary instabilities of the harmonic waves developing when the vibration amplitude gradually increases. From the dynamical system theory it is known that any time-periodic solution may lose its stability via three different bifurcations: the period-doubling, torus or a saddle-node. By fine-tuning the driving frequency or changing the viscosity of the fluid, in our study we establish that all three possibilities are possible for harmonic waves.

In the case of the supercitical period-doubling bifurcation, a sharp subharmonic peak appears in power spectrum of the temporal response of the drop. On the contrary, a torus bifurcation gives rise to modulation sidebands around the primary harmonic peak and spectral broadening.  This finding is confirmed by our experimental results for ethanol and canola oil drops subjected to $70$\,Hz vibrations, where we observed supercritical period-doubling bifurcation for more viscous canola oil and predominantly torus bifurcation for less viscous ethanol.

Our results should find practical applications in the emergent research directions of liquid optomechanics \cite{Dah16, Maa16, Kam16, Chi17, Mak19, Shk19} and hybrid metamaterial structures \cite{Lau17, Hao18, Mak18}. Thus far, optomechanical structures have mostly been implemented by using the solid-state technology \cite{Asp14} because modern electronic, photonic and phononic devices and circuits are based on solid-state platforms. However, liquid-state optomechanical systems may posses unique and practically useful characteristics that cannot be achieved in a solid-state configuration without the need of using high-power excitation signals. For example, this is the case of giant acoustic nonlinearities observed in gas bubbles and liquid droplets \cite{Tsa83, Rud06, Mak19} and a family of complex and intriguing nonlinear effects observed in hydrodynamics, turbulence, and atmosphere science -- the Akhmediev breathers \cite{Akh87}. Although large and increasing theoretical effort has been made to recreate and utilise these nonlinear phenomena in solid-state configurations \cite{Ree03, Ree03_1, Xio14, Xio16, Cao16, Lau17, Hao18}, considerable stiffness of solid-state structures requires impracticably high powers to access the nonlinearity. This is in stark contrast with fluids whose softness \cite{deGen} allows accessing their nonlinear properties with low power signals produced and controlled by modern photonics devices such as optical fibres and integrated-circuit resonators \cite{Dah16, Maa16, Kam16, Chi17, Mak18, Mak19, Shk19}.

Our results can also find applications in the emergent field of acoustic frequency comb generation \cite{Gan17} and their application in underwater distance measurements \cite{Wu19}. Similar to an optical frequency comb, an acoustic frequency comb is a spectrum consisting of a series of discrete, equally spaced elements that have a well-defined phase relationship between each other (for a review see, e.g., \cite{Mak19}). Optical frequency combs have typically been produced by mode-locked lasers or exploiting nonlinear optical effects in optical fibres and nonlinear photonic microresonators. Similarly, in Ref.~\onlinecite{Gan17} nonlinear acoustic effects in a solid-state device were used to generate a spectrum consisting of equally spaced and phase coherent comb lines. However, the effect observed in this present work can be used to generate frequency combs in a liquid-state system and potential in underwater settings, which should benefit marine sciences, underwater positioning and navigation \cite{Wu19}.     

\begin{acknowledgments}
This work was supported by the Australian Research Council (ARC) through its Future Fellowship (FT180100343). 
\end{acknowledgments}

\end{document}